\documentclass[aps,physrev,print,superscriptaddress]{revtex4-2}

\usepackage{graphicx}
\usepackage{dcolumn}
\usepackage{bm}
\usepackage[hidelinks]{hyperref}
\usepackage[mathlines]{lineno}
\linenumbers\relax
\usepackage{amssymb}
\usepackage{amsmath}
\nolinenumbers  
\usepackage{xcolor}
\usepackage[figure]{hypcap}
\usepackage[version=4]{mhchem}
\usepackage{subscript}
\usepackage{siunitx}
\DeclareSIUnit\Ions{ions}
\DeclareSIUnit\Ions{ions}
\DeclareSIUnit\electronvolt{eV}
\DeclareSIUnit\electrons{e^{-}}
\DeclareSIUnit\grooves{grooves}
\DeclareSIUnit\u{u}
\DeclareSIUnit\n{n}
\DeclareSIUnit\days{d}
\DeclareSIUnit\barn{b}
\usepackage{placeins}
\usepackage{chemmacros}

\begin{document}


\title{\textbf{Influence of carbon dioxide and water concentration on terbium thin films produced by Molecular Plating}}

\author{Ernst Artes}
\email[]{Contact author: erartes@uni-mainz.de}
\homepage[]{superheavies.de}
\affiliation{Department of Chemistry - TRIGA Site, Johannes Gutenberg University Mainz, 55128 Mainz, Germany}
\affiliation{Helmholtz Institute Mainz, 55128 Mainz, Germany}
\affiliation{GSI Helmholtzzentrum f{\"u}r Schwerionenforschung GmbH, 64291 Darmstadt, Germany}

\author{Primiana Cavallo}
\affiliation{Department of Chemistry - TRIGA Site, Johannes Gutenberg University Mainz, 55128 Mainz, Germany}

\author{Tobias H{\"a}ger}
\affiliation{Institut f{\"u}r Geowissenschaften, Johannes Gutenberg-Universit{\"a}t Mainz, Johann-Joachim-Becher-Weg 2, 55128 Mainz, Germany}

\author{Carl-Christian Meyer}
\affiliation{Department of Chemistry - TRIGA Site, Johannes Gutenberg University Mainz, 55128 Mainz, Germany}
\affiliation{Helmholtz Institute Mainz, 55128 Mainz, Germany}

\author{Christoph Mokry}
\affiliation{Department of Chemistry - TRIGA Site, Johannes Gutenberg University Mainz, 55128 Mainz, Germany}
\affiliation{Helmholtz Institute Mainz, 55128 Mainz, Germany}

\author{Dennis Renisch}
\affiliation{Department of Chemistry - TRIGA Site, Johannes Gutenberg University Mainz, 55128 Mainz, Germany}
\affiliation{Helmholtz Institute Mainz, 55128 Mainz, Germany}

\author{J{\"o}rg Runke}
\affiliation{Department of Chemistry - TRIGA Site, Johannes Gutenberg University Mainz, 55128 Mainz, Germany}
\affiliation{Helmholtz Institute Mainz, 55128 Mainz, Germany}
\affiliation{GSI Helmholtzzentrum f{\"u}r Schwerionenforschung GmbH, 64291 Darmstadt, Germany}

\author{Evgenia Schaffner}
\affiliation{Department of Chemistry - TRIGA Site, Johannes Gutenberg University Mainz, 55128 Mainz, Germany}

\author{Alice Seibert}
\affiliation{European Commission, Joint Research Centre (JRC) - Karlsruhe, Hermann-von-Helmholtz-Platz 1, 76344 Eggenstein-Leopoldshafen, Germany}

\author{Christina Trautmann}
\affiliation{GSI Helmholtzzentrum f{\"u}r Schwerionenforschung GmbH, 64291 Darmstadt, Germany}
\affiliation{Technische Universit{\"a}t Darmstadt - Materialwissenschaft, 64287 Darmstadt, Germany}

\author{Christoph E. D{\"u}llmann}
\affiliation{Department of Chemistry - TRIGA Site, Johannes Gutenberg University Mainz, 55099 Mainz, Germany}
\affiliation{Helmholtz Institute Mainz, 55099 Mainz, Germany}
\affiliation{GSI Helmholtzzentrum f{\"u}r Schwerionenforschung GmbH, 64291 Darmstadt, Germany}

\date{\today}

\begin{abstract}
Terbium and thulium thin films were produced by Molecular Plating under controlled conditions to elucidate a possible influence of water and carbon dioxide present in the plating solution. Platings were made in a glovebox with variable concentration of residual water and \ce{CO2} in a controlled inert atmosphere to study the impact on the quality of the produced thin films and on deposition yields. The morphology of the thin films was analyzed by scanning electron microscopy. The deposition yield was determined by neutron activation analysis at the research reactor TRIGA Mainz. Chemical analysis of the deposited layers was conducted using a combination of infrared, Raman and X-ray photoelectron spectroscopy. The Raman and IR spectra reveal the formation of hydroxides, oxides and carbonates. Water in the plating solution affects the quality of the thin films when its concentration exceeds 1 vol.-\%. The presence of \ce{CO2} leads to an increased carbonate content, which negatively influences the film quality. 

\begin{description}
\item[keywords] 
electrochemical deposition, infrared spectroscopy, Molecular Plating, Raman\\
spectroscopy, thin film deposition, X-ray photoelectron spectroscopy
\end{description}
\end{abstract}

\maketitle
\section{\label{sec:introduction}Introduction}
Electrochemical deposition of radionuclides has been researched for over 100 years \cite{Toedt1924}. Based on these studies Molecular Plating (MP) was developed 60 years ago by Parker and Falk \cite{Parker1962} for the production of thin films. \cite{Parker1962,Parker1964,Parker1964a,Getoff1967,Getoff1969, Eberhardt2008, Runke2013, Loveland2015, Loveland2009}. In MP the element of interest is electrochemically deposited from organic solutions such as alcohols \cite{Eberhardt2008}. This method is still used as a standard to produce thin films on various substrates like titanium or carbon for various nuclear physics applications \cite{Runke2013, Loveland2009, Loveland2015, Greene1999, Haas2020, Trautmann1989}, including targets used for ion beam experiments for the production of superheavy elements \cite{Runke2013, Duellmann2022, Duellmann2022a, Lommel2023}.\\
The production of superheavy elements necessitates the use of actinides such as $^{242,244}$Pu, $^{243}$Am, or $^{249}$Bk. The rarity of these isotopes necessitates the use of a technique that can achieve high deposition yields, which is possible with Molecular Plating. This technique enables the deposition of thin films with a thickness ranging up to about 1000~\si[per-mode=fraction]{\micro\gram\per\square\centi\meter} on substrates such as thin titanium foils with a thickness of 2.3~\si{\micro\meter}. It is imperative that the targets have a homogeneous surface and exhibit sufficient resilience to withstand the impact of heavy ions, such as $\approx5~\si[per-mode=fraction]{\mega\electronvolt\per\u}$ $^{48}$Ca, at ion fluences of 10$^{17}$~\si[per-mode=fraction]{\Ions\per\square\centi\meter}.
Despite the long time, for which MP has been employed, the deposition process and the chemical composition of the thin films are still being studied but yet not fully understood \cite{Vascon2012, Hansen1959}. It is, e.g., not fully clear why the quality and yield of the depositions often vary from deposition to deposition, despite identical experimental parameters.\\
Over the years, it has been found that the deposition is predominantly a basic precipitation of the element of interest \cite{Artes2023, Crespo2012, Hansen1959, Klemencic2010}. Spectroscopic studies suggested that organic components such as carbonates and carboxylates are present in the thin films alongside oxides and hydroxides \cite{Artes2023,Vascon2012}. Carbonates can be formed by the presence of \ce{CO2} and water in the reaction solution, by forming carbonic acid and its deprotonation. The concentration of \ce{CO2} and water in the solvents can vary with the age of the solvents and the atmospheric conditions under which they are stored and used by diffusion \cite{fogg2017carbon}.\\
Since targets are produced with rare isotopes, it is particularly undesirable if the deposition fails for no apparent reason. Therefore we have conducted a systematic investigation of the influence of \ce{CO2} and water on layers produced by the MP technique. Instead of rare actinides, terbium and thulium are used as lighter homologous and are deposited on titanium foil. Unlike the actinides, terbium and thulium are not radioactive, which simplifies handling and allows for characterization in laboratories, that are not certified for work with radioactive samples.\\

\section{Experimental section} 

    \subsection{Target production} 
\begin{figure}[ht!]
		\begin{center}
			\includegraphics[width=12cm]{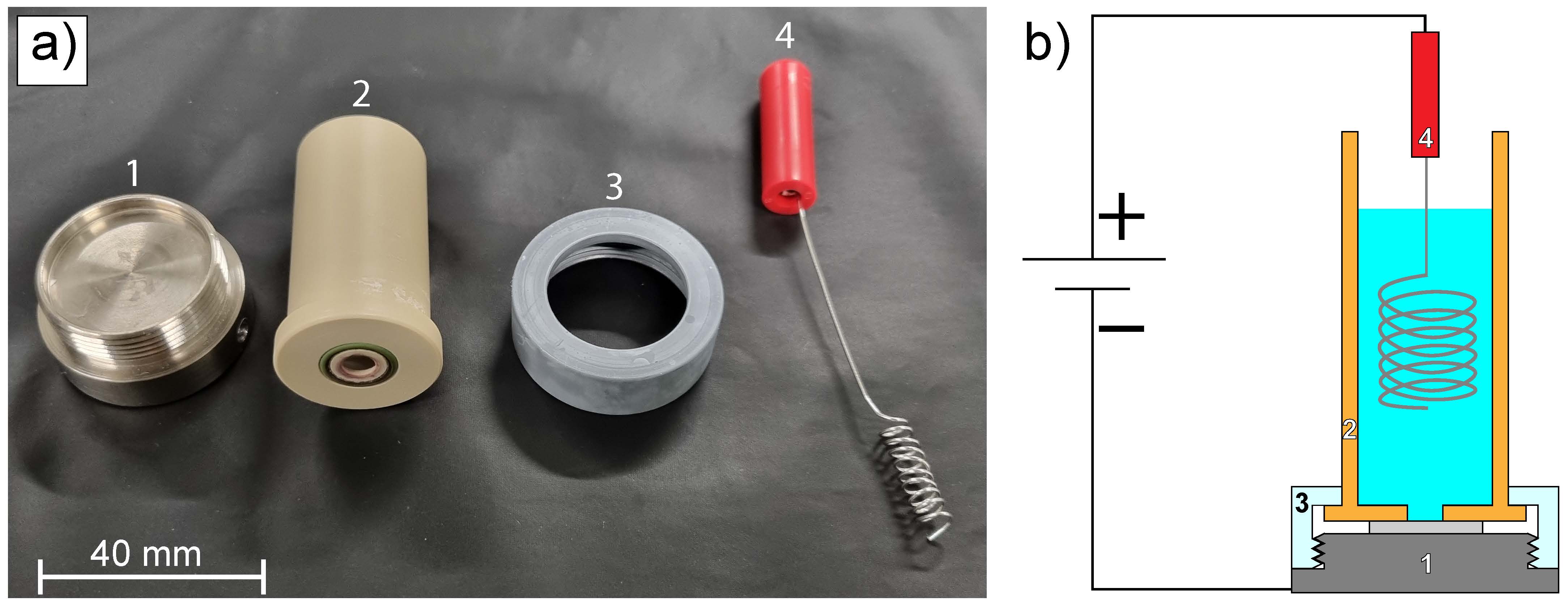}
			\caption{a) The single parts of the cell (from left to right) consist of a titanium base (1), the PEEK cell with gasket and deposition opening with a diameter of 6~\si{\milli\meter} (2), a nut to fix the cell (3) and the palladium anode (4). b) A scheme of the fully assembled and electrically connected PEEK cell is shown.}
			\label{Chimney cell}
		\end{center}	
	\end{figure}

{\label{taget production}}
    All targets were prepared using terbium(III)\,nitrate (\ce{Tb(NO3)3} CAS:~10043-27-3), or thulium(III)\,nitrate (\ce{Tm(NO3)3} CAS:~36548-87-5) in analytical quality. The used solvents isobutanol (IB, CAS:78-83-1, 2-methylpropan-1-ol), isopropanol (IP, CAS:67-63-0, propan-2-ol) had water contents of 30-50~ppm according to the supplier. \ce{D2O} (CAS:~7789-20-0) was used to add water to the reaction solution. All chemicals were purchased from Merck KGaA, Darmstadt, Germany. The reactions were prepared in an argon glovebox (MBraun) to keep ambient moisture and \ce{CO2} as low as possible. \ce{D2O} was used with the intention to perform $^{2}$H-NMR measurements of the supernatant solution. However, the NMR results showed no conclusive results.\\
    The targets were deposited on 10-\si{\micro\meter} thick titanium foil from an alcoholic solution containing terbium and thulium using an electrochemical cell by applying a constant current of 0.7~\si[per-mode=fraction]{\milli\ampere\per\square\centi\meter} for 2~\si{\hour}. The voltage between anode and cathode varied between 100 and 600~\si{\volt} for the depositions.\\
    The molecular plating was performed in a small cell with a volume of 10~\si{\milli\liter}, which is shown in Fig. \ref{Chimney cell}. It features a circular deposition area with a diameter of 6~\si{\milli\metre}. The substrate for the deposition was a titanium foil with a thickness of 10~\si{\micro\meter} and a purity of 99.6~\%. The titanium foil was purchased from HMW Hauner GmbH \& Co. KG. A palladium wire (1~\si{\milli\meter} diameter) served as an anode. It was curled at the end to increase the surface toward the cathode.
	
    The solvent used was a mixture of isobutanol (IB) and isopropanol (IP) in a ratio 9:1. Terbium nitrate and thulium nitrate were each dissolved in the solvent to get stock solutions with concentrations of 7.1~\si[per-mode=fraction]{\micro\gram\per\micro\liter}. \ce{CO2} was passed through the IB/IP~=~9:1 mixture for 15~\si{\minute} to obtain a \ce{CO2}-saturated solution, which features a \ce{CO2}-concentration of 0.18~\si[per-mode=fraction]{\mole\per\liter} at $20^\circ$C for this specific mixture \cite{fogg2017carbon}.\\
    The saturated solution was mixed with a \ce{CO2}-free mixture containing the lanthanide salt. In this way, final solutions with different saturation levels of \ce{CO2} (depending on the ratio between \ce{CO2}-saturated and untreated solvent) were produced. The untreated solvent was considered as \ce{CO2}-free as procured.\\
    An aliquot of the stock solution containing 142~\si{\micro\gram} of the lanthanide was mixed with the \ce{CO2}-saturated alcoholic mixture to set the \ce{CO2}-concentration in the solution. As a result, each deposition led to an areal weight of 500~\si[per-mode=fraction]{\micro\gram\per\square\centi\meter} by assuming quantitative deposition.\\
    Afterwards \ce{D2O} was added to that solution, to obtain the desired \ce{D2O}-concentration. The volume was adjusted to 10~$\si{\milli\liter}$ with \ce{CO2}-free alcoholic mixture to achieve the desired concentration of \ce{CO2} and \ce{D2O}. Targets were prepared with \ce{D2O}-concentrations of 0, 0.5, 1.0, 1.5 and 2.5 vol.-\% and \ce{CO2}-saturations of 0, 20, 40 60, 80 and 100~\%. By varying both parameters fourteen targets were produced, as displayed in a matrix (table \ref{Morphology TbNO3}) that encompasses both parameters.

	\subsection{Target analysis}
    Different methods were used to analyze the samples. Samples prepared with the salt \ce{Tb(NO3)3} were analyzed in a scanning electron microscope (SEM), and infrared (IR) and Raman spectra were recorded. The yields of the depositions were determined indirectly by neutron activation analysis (NAA) by determining the lanthanide concentration left in the supernatant solution after deposition \cite{Liebe2008}.\\
    X-ray photoelectron spectroscopy (XPS) was performed on samples prepared with \ce{Tm(NO3)3}.

        \subsubsection{Neutron activation analysis}
		Neutron activation analysis (NAA) is a simple and sensitive method of determining the quantity of a substance \cite{Eberhardt2019}. The terbium deposition yields were determined by NAA. For this purpose, aliquots of the solutions were taken before and after deposition and activated in the research reactor TRIGA Mainz at a neutron flux of $\num{7e+11}$~\si[per-mode=fraction]{\n\per\square\centi\meter\per\second} for 1~\si{\hour}. Terbium is a monoisotopic element ($^{159}$Tb). This isotope has a neutron capture cross section of 23.6~\si{\barn}. The neutron capture product $^{160}$Tb has a half-life of 72.3~\si{\days} and undergoes $\beta^{-}$ decay (100~\%), which is followed by a gamma-ray emission [$E_{\gamma}$ = 879~\si{\kilo\electronvolt} (30.1~\%); 299~\si{\kilo\electronvolt} (26.1~\%) and 966~\si{\kilo\electronvolt} (25.1~\%)] \cite{Nica2021}. After neutron irradiation, the gamma activities of the samples were measured in a high purity germanium (HPGe) detector (Ortec) and the gamma spectra were evaluated with the Genie 2000 software (Canberra). From the difference of the activities of the samples before and after deposition, the deposition yield, is obtained, assuming that all of the lanthanides no longer present after the plating procedure were deposited on the substrate. The thin films were not measured directly as their geometry precluded their activation due to the fact that bending the substrate in order to irradiate it in the reactor would destroy the thin film.

        \subsubsection{Scanning electron microscopy}
        To evaluate the surface morphology of the films, the samples were examined in a scanning electron microscope (SEM, Philips XL30) with an accelerating voltage of 20~\si{\kilo\volt} in combination with detectors for secondary (SE) and backscattered electrons (BSE).
	
        \subsubsection{Confocal Raman spectroscopy}	
        The targets were analyzed with a confocal Raman microscope (HR800 system, Horiba Jobin Yvon) of the Materials Research Department at GSI Darmstadt. A laser with 473~\si{\nano\meter} emission wavelength, a grating with 1800~\si[per-mode=fraction]{\grooves\per\mm}, and a slit width of 100~\si{\milli\meter} were used to acquire the data. The spectral resolution was 0.8~\si{\per\centi\meter}. The acquisition time for one spectrum was set to 20 \si{\s} for all measurements. Raman spectra of the different terbium nitrate samples from \SIrange{200}{3200}{\per\centi\meter} were collected.
	 
        \subsubsection{Infrared microscopy}
        The targets were analyzed by infrared microscopy (Nicolet Continu$\mu$m connected to a Nicolet 6700, THERMO SCIENTIFIC) at the Institute of Geosciences of the Johannes Gutenberg University Mainz. A mercury cadmium telluride (short: MCT) detector was used. The increments were set to 0.1~\si{\nano\metre}. The acquisition time for each spectrum was set to 5~\si{\min}. Infrared spectra from \SIrange{650}{4000}{\per\centi\meter} were collected in reflection on a gold mirror. IR spectra were recorded from three different locations of each terbium thin film produced to exclude artifacts in the film. The reflection of IR radiation from the sample was measured. Initially, the substrate, which was titanium, was measured for background correction. The Ti background was subtracted from the recorded spectra.
	
        \subsubsection{X-ray photoelectron spectroscopy}
        The XPS spectra were recorded at the Joint Research Center (JRC) in Karlsruhe, Germany \cite{Gouder2019}. For the XPS analysis of the \ce{D2O}/\ce{CO2} matrix samples, the thulium targets were used because the relevant XPS peaks in the investigated thulium and oxygen binding energy regions are not superimposed or interefered by other contributions. Targets from 0.5 and 1~vol.-\% \ce{D2O}-concentration and \ce{CO2}-saturation from 0 to 80~\% were producedas described in section \ref{taget production} and placed on designated XPS sample holders. These targets were enclosed in an airtight container and transported to JRC Karlsruhe where they were introduced into the ultra-high vacuum (UHV) chamber for the XPS measurements via a two-stage loading system. The facility at JRC Karlsruhe is described in \cite{Gouder2019}.\\
        The XPS measurements were conducted several weeks after sample preparation and days after introduction to the UHV system. They were also stored during the measurement campaign of 2 weeks in a chamber with a base pressure of about $2.0\cdot10^{-10}~\si{\milli\bar}$ adjacent to the XPS analysis chamber. XPS measurements were performed using a monochromatic Al K$_{\alpha}$-X-ray source (XRC-1000 MF) equipped with a $\mu$-FOCUS 500 monochromator and a Phoibos 150 hemispherical analyzer controlled by a HAS 3500 plus system (both from Specs, Berlin). The spectrometer was calibrated using the Au metal's 4\textit{f}$_{7/2}$ (83.9(1)~\si{\electronvolt}) and Cu metal's 2\textit{p}$_{3/2}$ (932.7(1)~\si{\electronvolt}) reference lines. Surveys and high-resolution spectra were recorded at pass energies of 50 or 20~\si{\electronvolt}. To suppress sample charging, a flood gun FG-500 (Specs, Berlin) was available. All measurements were conducted at room temperature using an in-house data acquisition program.\\

\section{Results and discussion}

    \subsection{Morphology}
    The morphology of the samples prepared from \ce{Tb(NO)3}, examined with the scanning electron microscope and a conventional USB optical microscope (Traveler), are shown in table \ref{Morphology TbNO3}, tabulated according to the \ce{D2O} and \ce{CO2}-concentration.\\
    The images of the thin films show that a deposition without the presence of \ce{D2O} or \ce{CO2} is not possible. The light microscope image shows a slightly darker shade of the titanium substrate where the deposition took place. However, the SEM image clearly shows the substrate and the rolling marks from production. There is no significant height contrast, indicating that the surface is relatively smooth.  Addition of \ce{CO2} to the solution facilitates deposition, but the thin films show only loose tiles with wide cracks in between, as seen in the SEM images shown in Tab. \ref{Morphology TbNO3}. This trend is even more pronounced when the \ce{CO2} content in the solution is high. Clearly a deposition can be seen at a water concentration of 0.5 - 1~vol.-\%. This is visible due to the white layer on the titanium substrate. Additionally, in the SEM pictures there is clearly a layer visible, which shows "mudcracking", which is already described for molecular plated targets by e.g. A. Vascon et al. \cite{Vascon2012}. If the \ce{CO2} content of the solution increases at these water concentrations, the thin films become brittle and the mudcracking becomes more intense. At 80~\% \ce{CO2} saturation, the tiles are loose and the thin film does no longer adhere well to the substrate. This is visible with the naked eye and especially in the corresponding SEM picture for the sample with 0.5~vol.-\% \ce{D2O}.\\
    It can be seen that the film quality again decreases dramatically, if the water concentration in the reaction solution exceeds 1~vol-\%. The layer is inhomogeneous and a look at the surface with the naked eye shows a brittle film. In the SEM image the films do no longer show the typical mudcracking, but rather singular regions with granular material without significant adherence to the substrate.\\
    Table \ref{Morphology TbNO3} highlight the crucial role of \ce{D2O} and \ce{CO2} levels on the film quality and yield. Molecular Plating needs the presence of water in the solution. \ce{D2O} concentration of 0.5 or 1 vol.-\% facilitates Molecular Plating, while higher water content results in weak adhesion and more cracked films, as exemplified by the sample with 1.5 vol.-\%. To achieve optimal results, it is recommended to have a water content of 0.5 to 1 vol.-\% in the solution. The presence of water in Molecular Plating induces cathodic reduction, converting water into hydrogen and hydroxide ions. The hydroxide ions create a local alkaline environment at the cathode, promoting deposition as is long known \cite{Hansen1959}. However, we assume that the generation of hydrogen has negative effects on the film quality. An increase of water content triggers excessive hydrogen formation. This leads to a constant production of hydrogen at the surface hindering the formation of a thin layer at this particular point, thereby preventing adhesion, leading to the formation of a heavily cracked film, and to reduced adhesion and homogeneity.\\

\begin{table}[]
\centering
\caption{Morphology and yield (given as percentage value in the corresponding field) of terbium thin films produced by Molecular Plating with varying water and \ce{CO2}-saturation. For selected samples, a picture from a conventional microscope is shown on the left side (diameter 6~\si{\milli\meter}. On the right a SEM picture of the sample with the size of 130~x~130~\si{\micro\meter} is shown for some samples.}
\label{Morphology TbNO3}
{\includegraphics[width=11.5cm]{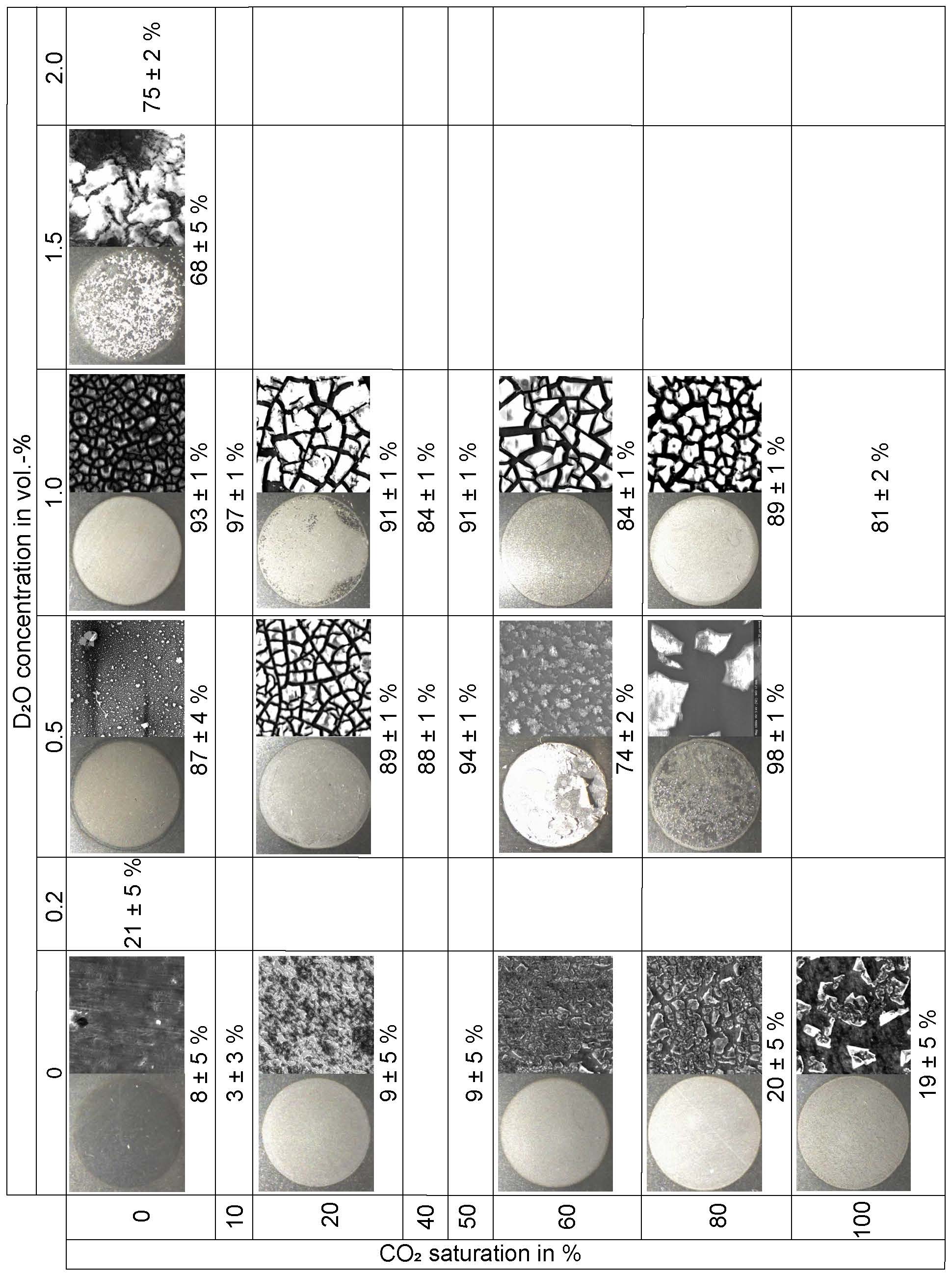}}
\end{table}

    \subsection{Yields}
    Deposition yields of terbium were indirectly determined by NAA of the supernatant solution. The yields as function of the \ce{D2O} and \ce{CO2}-concentrations are given in Tab. \ref{Morphology TbNO3}.\\
    The uncertainties are based on the net number of counts of the gamma spectra of the supernatant solution before and after deposition, and the resulting Gaussian error propagation.\\
    Without any addition of water beyond the amounts already present in the solvents the deposition yield remains below 20~\% and reaches such levels only at elevated \ce{CO2} concentrations. In a water concentration of 0.5-1.0 vol.-\%, the yield increases significantly, reaching levels of 90~\%, and does become independent of the \ce{CO2} concentration. However, higher \ce{CO2} concentrations under these conditions result in films that are more brittle and show poor adhesion to the substrate. At yet higher water concentrations, the yield decreases again. The influence of \ce{CO2} in this regime was not studied, as the negative effects associated with increasing \ce{CO2} concentrations had already been observed at lower water concentrations.\\
    Even trace amounts of water, in the order of a few ppm as already present in the original solution, appear sufficient to promote the formation and precipitation of carbonates, albeit this results in small yields (<20~\%) and films with poor adhesion to the substrate.\\
    Molecular Plating can produce thin films even without the addition of water, if the \ce{CO2} concentration is high enough. With small amounts of water in the range of a few ppm already present in the solution, carbonates can be formed and precipitated in the process. However, the adhesion to the substrate is weak and the deposition yield does not exceed 20~\%. The presence of \ce{CO2} might alter the film composition by incorporating carbonates and formates, facilitating deposition. However, this modified composition does lead to insufficient adhesion and low yields.\\
    Additionally, the introduction of \ce{CO2} to the reaction solution containing 0.5 or 1 \% \ce{D2O} results in diminished adhesion and film uniformity compared to films prepared without \ce{CO2}. SEM images reveal that cracks in the films become more pronounced as the \ce{CO2} concentration increases in the solution.\\

    \subsection{Raman and IR spectroscopy}

\begin{figure}[]
	\begin{center}
		\includegraphics[width=11.5cm]{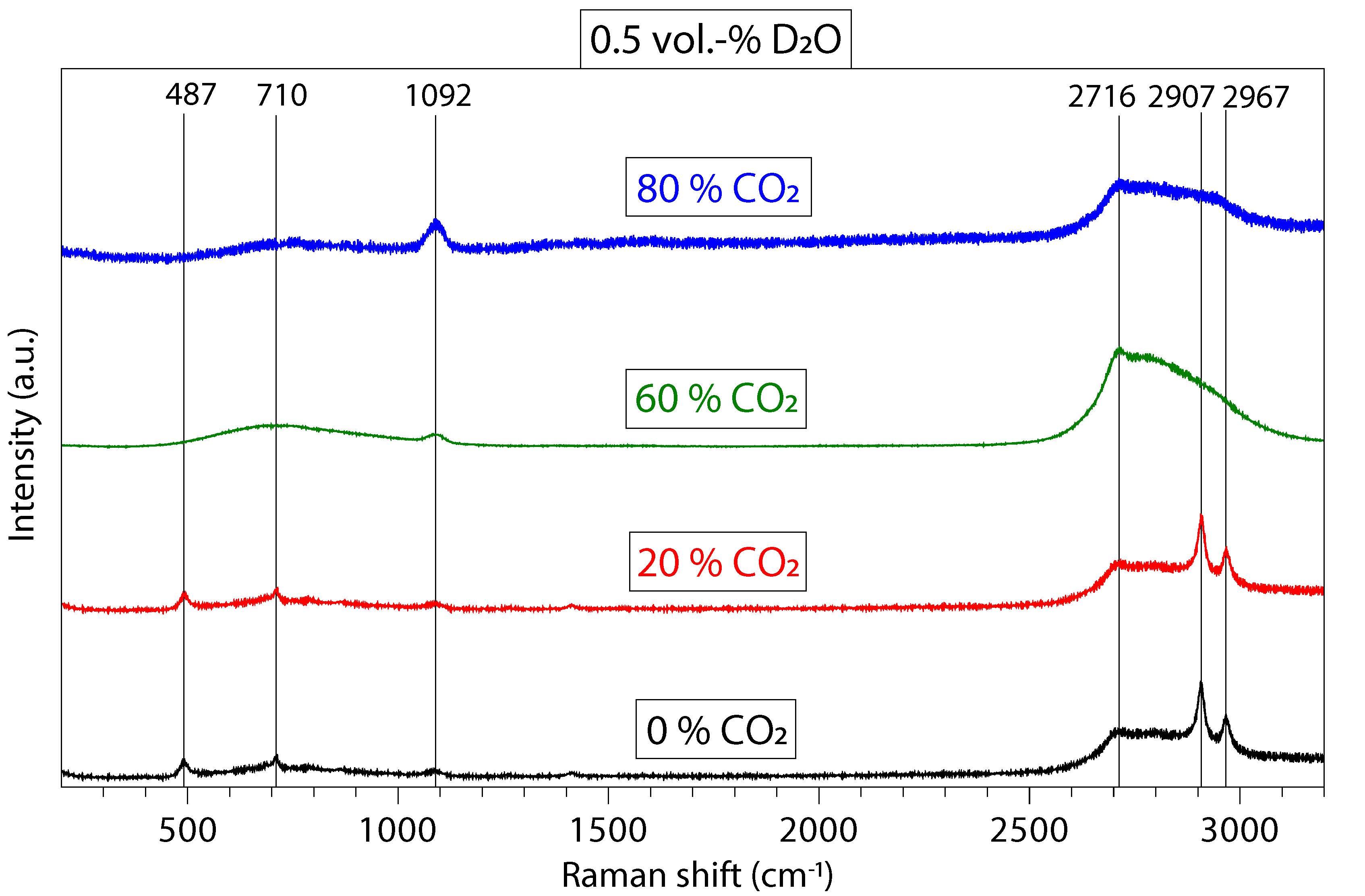}
		\caption{Stacked Raman spectra of terbium thin films produced by MP from solution with fixed water content of 0.5 vol\% \ce{D2O} and different \ce{CO2} saturation. All spectra are normalized to their respective maximum value. A detailed assignments of the vibrational bands are given in Tab.~\ref{vibration bands}.}
		\label{Raman 50}
	\end{center}	
	\end{figure}

\begin{figure}[]
	\begin{center}
		\includegraphics[width=11.5cm]{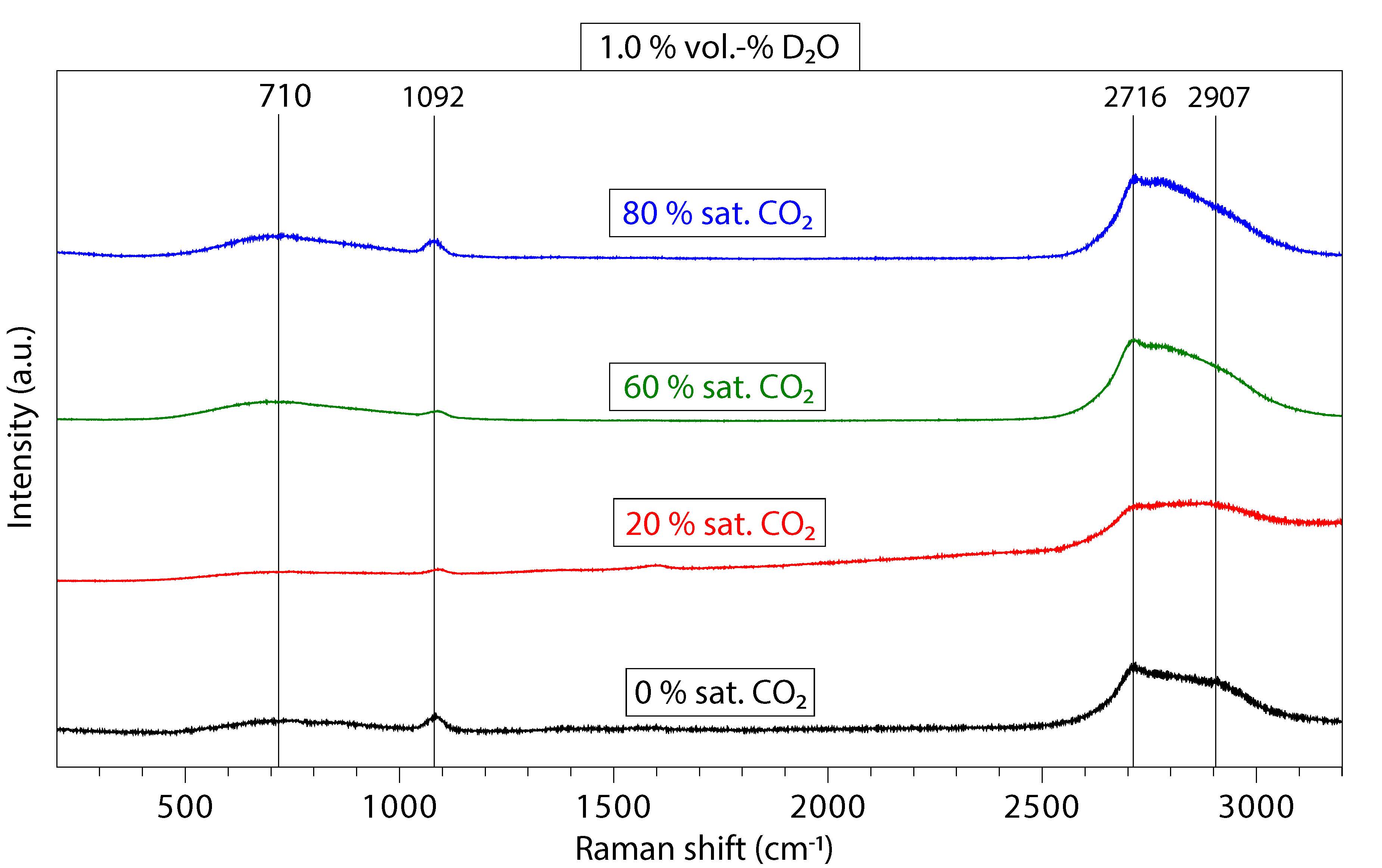}
		\caption{Stacked Raman spectra of terbium thin films produced by MP from solution with fixed water content of 1.0 vol\% \ce{D2O} and different \ce{CO2} saturation. All spectra are normalized to their respective maximum value. A detailed assignments of the vibrational bands are given in Tab.~\ref{vibration bands}.}
		\label{Raman 100}
	\end{center}	
	\end{figure}

        Spectra taken with the Raman microscope of all samples with 0.5~vol.-\% \ce{D2O} are shown in Fig. \ref{Raman 50} and those of all samples with 1.0~vol.-\% \ce{D2O} in Fig. \ref{Raman 100}. Due to the inhomogeneity and the mudcracking of the surface of the films, it is challenging to make quantitative statements. Consequently, qualitative statements are made about the individual spectra. The individual bands are then compared to the baseline of the spectra, thus enabling statements to be made about the influence of \ce{CO2} and \ce{D2O} on the samples.\\
        The observed vibrational bands and their assignments to the corresponding species \cite{Frost2007,Kartha1981,Silva2009,Spiridigliozzi2022,Sohn2014} are summarized in table \ref{vibration bands}. \\
        The Raman spectra of the samples with 0.5 vol.-\% of water show bands at approximately 500~\si{\per\centi\meter} (assigned to oxidic species), at 700~\si{\per\centi\meter} and 1100~\si{\per\centi\meter} (assigned to carbonate vibrations), and at 2700 - 3000~\si{\per\centi\meter} (assigned to formate vibrations). Notably, no formates or formic acid were intentionally added to the solutions.\\
        With increasing \ce{CO2}-concentrations, the oxidic vibration (500~\si{\per\centi\meter}) diminishes, while the carbonate vibrations (700~\si{\per\centi\meter} and 1100~\si{\per\centi\meter})  become more pronounced compared to the baseline. Additionally, the peak at 700~\si{\per\centi\meter} becomes significantly broader. The formate peaks also broadens, but does not intensify as much as the carbonate peak with increasing \ce{CO2}-concentrations. Frost et al. \cite{Frost2007} reported a weak carbonate band at the range of 1400 - 1500~\si{\per\centi\meter}, which could not be observed. Since its intensity is weak, we assume, that the band disappears due to the mudcracking and amorphicity of the sample.\\

\begin{figure}[]
	\begin{center}
		\includegraphics[width=11.5cm]{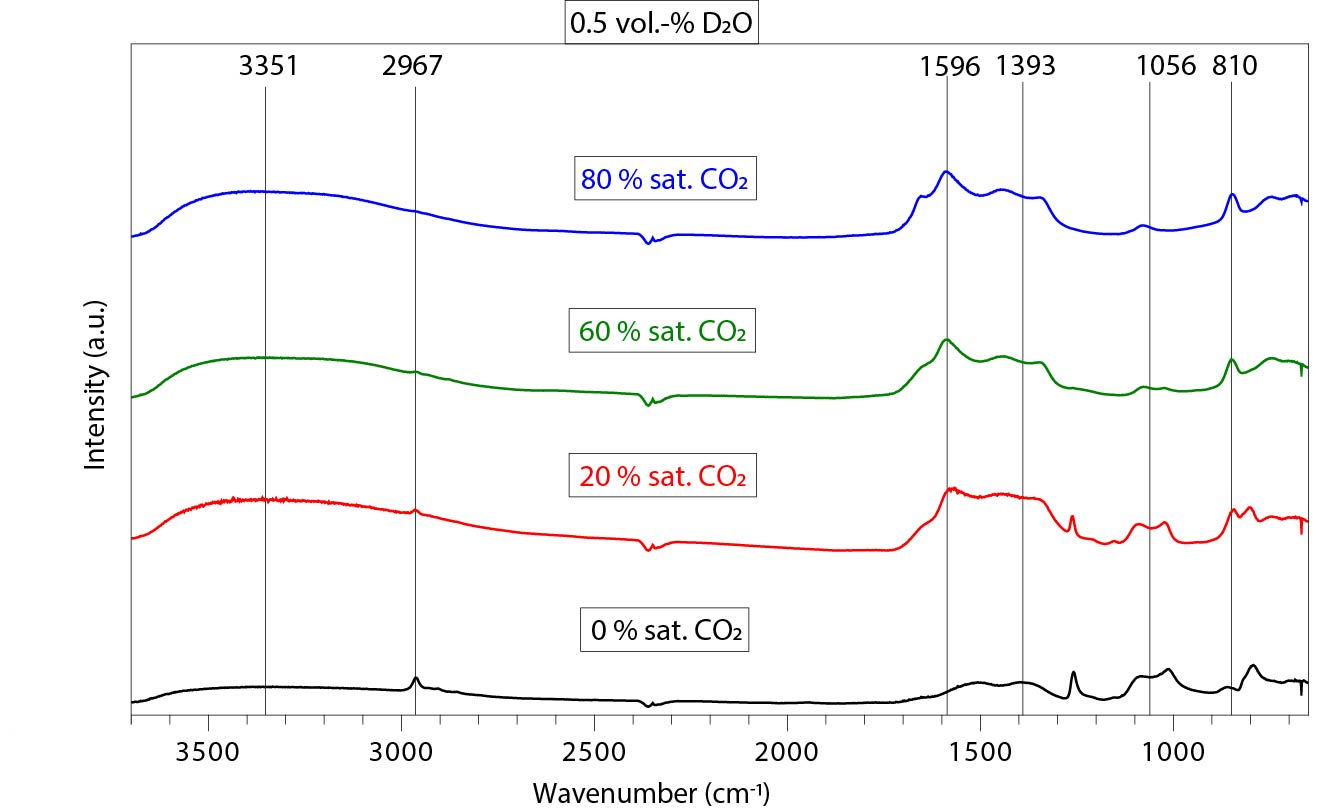}
		\caption{Stacked IR spectra of terbium thin films produced by MP from solutions with fixed water concentration of 0.5 vol-\% \ce{D2O} and different \ce{CO2} saturation. All spectra are normalized to their respective maximum value. A detailed assignments of the vibrational bands are given in Tab.~\ref{vibration bands}.}
		\label{IR 50}
	\end{center}	
	\end{figure}

\begin{figure}[]
	\begin{center}
		\includegraphics[width=11.5cm]{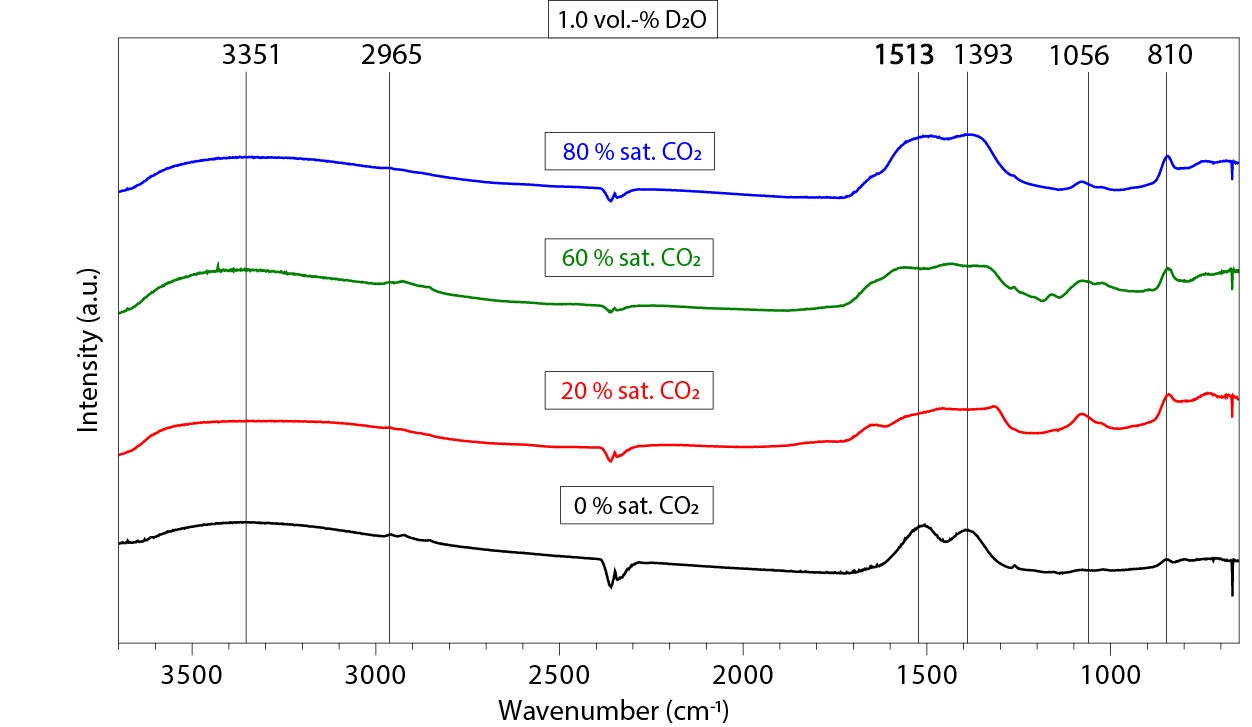}
		\caption{Stacked IR spectra of terbium thin films produced by MP from solutions with fixed water concentration of 1.0 vol-\% \ce{D2O} and different \ce{CO2} saturation. All spectra are normalized to their respective maximum value. A detailed assignments of the vibrational bands are given in Tab.~\ref{vibration bands}.}
		\label{IR 100}
	\end{center}	
	\end{figure}

        This broadening phenomenon of vibrational bands is observed consistently in all spectra of samples with 1 vol.-\% water concentration.\\
        With increasing water content in the samples, it should be assumed that the oxide bands are more pronounced. However, these were no longer observed in the Raman spectra. We suspect that the oxidic bands are still present, but are no longer observable in the Raman spectra, because more cracks appear in the thin films. These cracks enhance unintentional light scattering, thereby diminishing the intensity of the detected Raman bands.\\
        Spectra taken with the infrared microscope are shown in Fig. \ref{IR 50} and Fig. \ref{IR 100}. Like for the Raman spectra, all samples with 0.5 vol.-\% and with 1.0 vol.-\% \ce{D2O} were combined in one graph each. In table \ref{vibration bands} all bands and assigned species \cite{Silva2009,Spiridigliozzi2022,Kartha1981,Caro1972,Sohn2014} are given. As described above, qualitative statements are also predominantly made here. Here too, the intensities of the individual bands compared to the baseline are taken into account for the evaluation.\\
        The infrared spectra reveal the presence of oxides, carbonates, and also formates. A broad band at 3600 - 2800 \si{\per\centi\meter} indicates hydroxide vibrations. At 2967 and 2965~\si{\per\centi\meter}, a formate band is evident, but it diminishes with increasing \ce{CO2}-content. At 2400~\si{\per\centi\meter}, a vibrational band associated with titanium is observed in the spectra. \\
        At 1596~\si{\per\centi\meter}, carbonate, and formate bands overlap, with additional overlap at 1393~\si{\per\centi\meter} and 1056~\si{\per\centi\meter}, making distinct identification challenging. The only interference free bands of an organic compound is the formate band at 2950~\si{\per\centi\meter}. Another carbonate band can be observed at 810~\si{\per\centi\meter}.\\
        As the concentrations of water and \ce{CO2} increase, a broadening of the carbonate bands is evident.\\

\begin{table}[]
\centering
\caption{The assignment of the Raman and IR bands to the corresponding vibrations and species.}
\label{vibration bands}
\begin{tabular}{ccccc}
\textbf{\begin{tabular}[c]{@{}c@{}}Chemical\\ species\end{tabular}} & \textbf{\begin{tabular}[c]{@{}c@{}}Vibration\\ mode\end{tabular}} & \textbf{\begin{tabular}[c]{@{}c@{}}Raman\\ band\\ \textbf{[cm$^{-1}$]}\end{tabular}} & \textbf{\begin{tabular}[c]{@{}c@{}}IR\\ band\\ \textbf{[cm$^{-1}$]}\end{tabular}} & \textbf{Reference} \\ \hline
\begin{tabular}[c]{@{}c@{}}hydroxide\\ residue water\end{tabular} & $\nu$ & - & 3351 & \cite{Klevtsov1967} \\
 &  &  &  &  \\
oxide  & $\nu_{4}$ & 487 & - & \cite{White1972} \\
 &  &  &  &  \\
carbonate & \begin{tabular}[c]{@{}c@{}}$\nu_{3}$\\ $\nu_{3}$\\ $\nu_{1}$\\ $\nu_{2}$\\ $\nu_{4}$\end{tabular} & \begin{tabular}[c]{@{}c@{}}-\\ -\\ 1092\\ -\\ 710\end{tabular} & \begin{tabular}[c]{@{}c@{}}1596\\ 1393\\ -\\ 1056\\ 810\end{tabular} & \begin{tabular}[c]{@{}c@{}}\cite{Frost2007, Spiridigliozzi2022, Frost_2004, Frost2013, RahimiNasrabadi2017}\end{tabular} \\
 &  &  &  &  \\
formate & \begin{tabular}[c]{@{}c@{}}$\nu_{1}$\\ $\nu_{1}$\\ $\nu_{1}$\\ $\nu_{5}$\\ $\nu_{2}$\\ $\nu_{6}$\end{tabular} & \begin{tabular}[c]{@{}c@{}}2967\\ 2907\\ 2716\\ -\\ -\\ -\end{tabular} & \begin{tabular}[c]{@{}c@{}}2965\\ -\\ -\\ 1596\\ 1393\\ 1056\end{tabular} & \cite{Frost2013, Kartha1981, Saralidze1967, Hosson1975} \\ \hline
\end{tabular}
\end{table}

\begin{figure}[]
	\begin{center}
		\includegraphics[width=11.5cm]{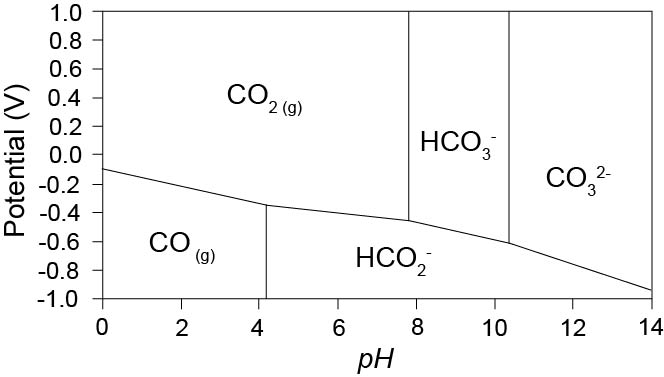}
		\caption{The Pourbaix diagram for \ce{CO2} depicts how \ce{CO2} reacts in water across varying pH levels and diverse potentials relative to a standard reference electrode. \cite{Koenig2019}}
		\label{Pourbaix}
	\end{center}	
	\end{figure}

        In addition to hydroxides and oxides, vibrational modes of carbonates and formates are visible in the spectra. The Raman and IR spectra reveal that formates are produced when the MP process occurs. This can be understood by the \ce{CO2} exposure to negative potentials and high pH value at the cathode. In this situation, \ce{CO2} can be transformed into formates through reduction, as indicated by the Pourbaix diagram shown in Fig. \ref{Pourbaix} \cite{Koenig2019}.\\
        However, the Raman spectra show that a significant portion of \ce{CO2} transforms into carbonates, leading to their precipitation. This is visible by the increased intensity of the carbonate bands in the Raman spectra at 710, 1092 and 2716~\si{\per\centi\meter}, while the formate peaks do not increase in intensity with increasing \ce{CO2} compared to the base line of each spectrum. The deposition of carbonates can occur immediately; however, the electrochemical production of formates at the cathode is a prerequisite for their subsequent deposition. The formation of formates at the cathode precedes the deposition process, which is the rate-limiting step in formate deposition.\\
        If the \ce{CO2} concentration increases in the solution, the amount of carbonates in the thin film also increases. In contrast the relative amount of oxides decreases. This shows that the formation of carbonates competes with the formation of hydroxides and oxides.\\

        \subsection{X-ray photoelectron spectroscopy}	
\begin{figure}[]
	\begin{center}
		\includegraphics[width=11.5cm]{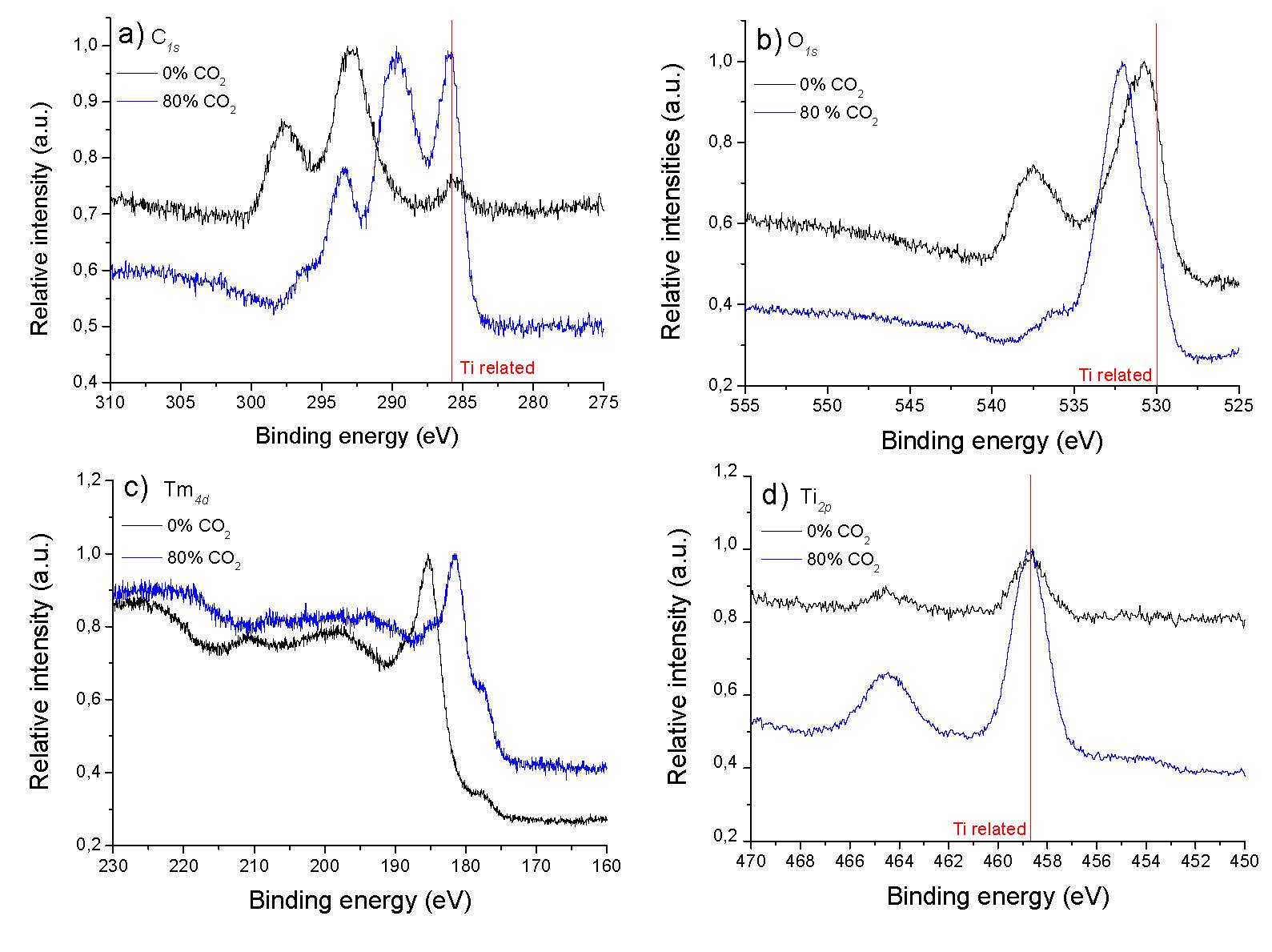}
		\caption{XPS spectra of C$_{1s}$, O$_{1s}$, Tm$_{4d}$ and Ti$_{2p}$ of samples with 0.5 vol-\% \ce{D2O} at 0 \% \ce{CO2} saturation (black) versus 80 \% \ce{CO2} saturation (blue). The red vertical lines represent species related to the titanium substrate.}
		\label{XPS05}
	\end{center}	
	\end{figure}

\begin{figure}[]
	\begin{center}
		\includegraphics[width=11.5cm]{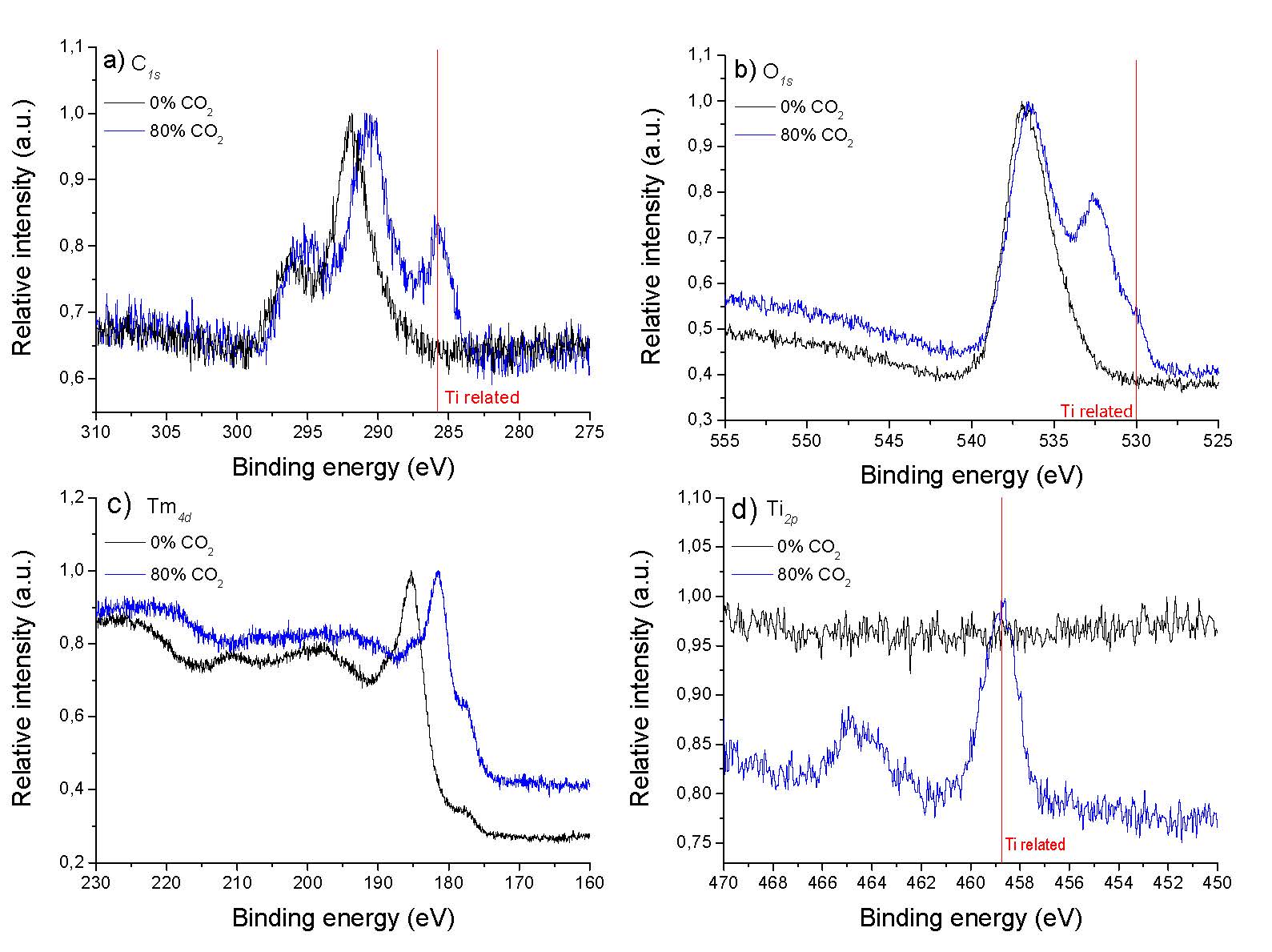}
		\caption{XPS spectra of C$_{1s}$, O$_{1s}$, Tm$_{4d}$ and Ti$_{2p}$ of samples with 1 vol-\% \ce{D2O} at 0 \% \ce{CO2} saturation (black) versus 80 \% \ce{CO2} saturation (blue). The red vertical lines represent species related to the titanium substrate.}
		\label{XPS1}
	\end{center}	
	\end{figure}

\begin{figure}[]
	\begin{center}
		\includegraphics[width=11.5cm]{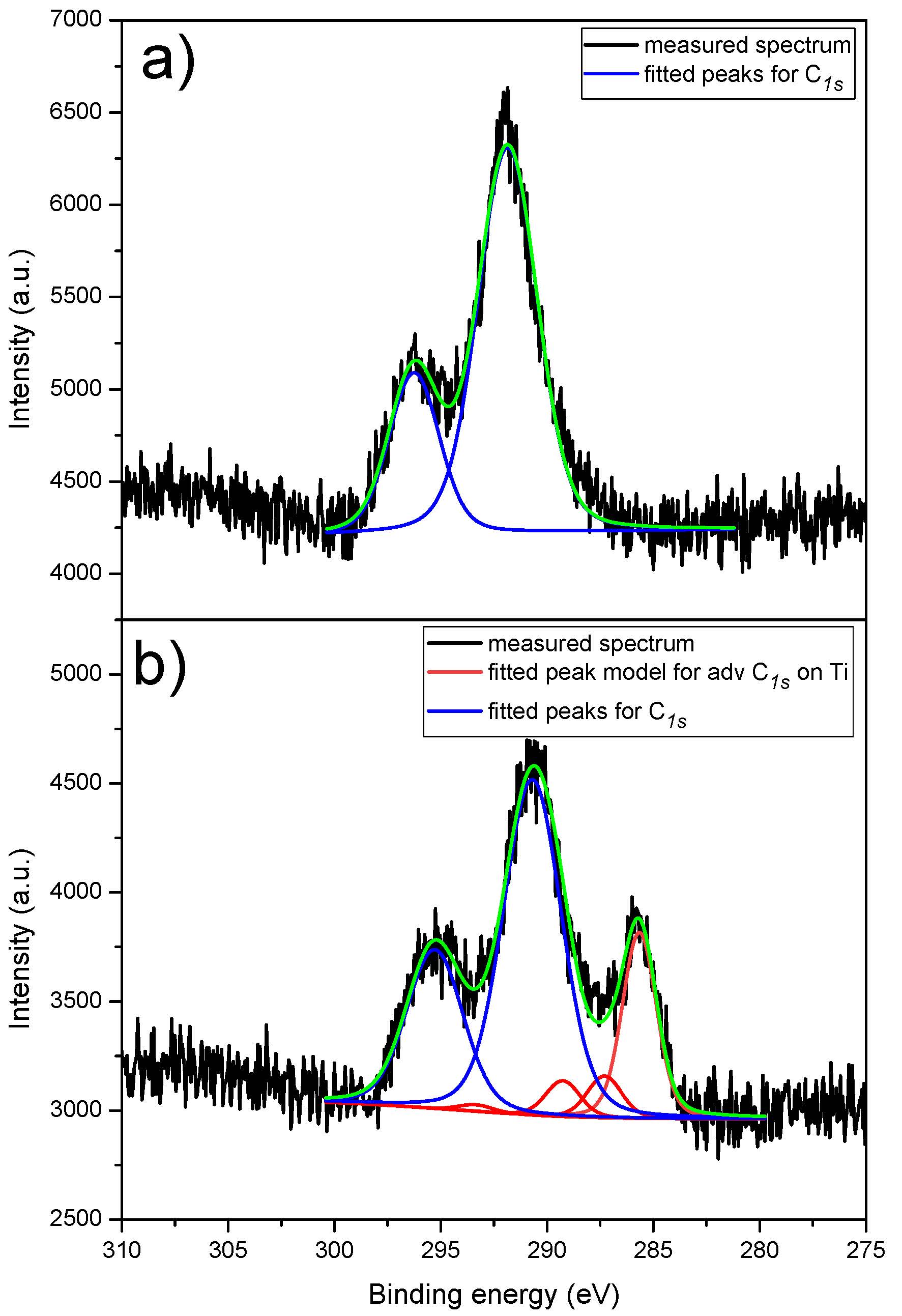}
		\caption{Comparison of C$_{1s}$ spectrum for a deposition from 1 vol-\% water and 0 \% \ce{CO2} (panel a) and a deposition from 1 vol-\% water and 80 \% \ce{CO2} (panel b). The black curves are the measured spectra. The blue curves are the fitted peaks for C$_{1s}$. The red curves are the fitted peak model for adventitious C$_{1s}$ on titanium. The green curve is the fitted envelope.}
		\label{C1s_fit}
	\end{center}	
	\end{figure}

Figures \ref{XPS05} and \ref{XPS1} display XPS detail spectra (C$_{1s}$, O$_{1s}$, Tm$_{4d}$, Ti$_{2p}$) for samples with 0.5 and 1 vol-\% water at 0 and 80~\% \ce{CO2} saturation. The thick thulium deposits and use of a monochromatic X-ray source caused charging, leading to shifted peak positions due to the insulating nature of the material. This could only be partially compensated by a flood gun.\\
Cracking and flaking of the thin films, similar to the terbium samples shown in table \ref{Morphology TbNO3}, expose the titanium backing, complicating spectral interpretation. Signals stemming from the titanium substrate appear in the C$_{1s}$ and O$_{1s}$ spectra in addition to the signals related to the deposited thulium layers. The Ti$_{2p}$ doublet at $\approx$458.7 and 464.4~\si{\electronvolt}, with 5.7~\si{\electronvolt} spin-orbit splitting, indicates a \ce{TiO2} surface layer, which is also present on untreated titanium foils. The titanium substrate related signals remain unshifted, indicating effective charge compensated.\\
Charging effects and overlapping signals hinder the precise assignment of the C$_{1s}$ and O$_{1s}$ peaks. A consistent C$_{1s}$ peak at $\approx$285~\si{\electronvolt}, which is highlighted by the red line in figure \ref{XPS05}a and \ref{XPS1}a correlates with the presence of Ti$_{2p}$ suggesting it originates from adventitious carbon on the titanium backing. A peak model for this C$_{1s}$ contribution was developed (Fig. \ref{C1s_fit}b, red curves) based on a titanium blank sample. Additional broad C$_{1s}$ peaks are observed and attributed to charge-shifted carbon species from the deposition.\\
Figure \ref{C1s_fit} compares C$_{1s}$ spectra from 1 vol-\% \ce{D2O} samples at 0~\% (panel a) and 80 \% \ce{CO2} (panel b). The fitting approaches shows beside the 285.7~\si{\electronvolt} peak related to titanium backing  broader, shifted components at around 290~\si{\electronvolt} and 296~\si{\electronvolt} attributed to carbon originating from the deposition.\\
The $\approx$4.5~\si{\electronvolt} peak separation and intensity ratio of $\approx$2.5 differ from typical adventitious carbon, suggesting higher-oxidized species.\\
Barr et al. \cite{Barr1995} identified C-C/C-H as the main component of adventitious carbon, with minor contributions from C-O and O-C=O (as seen for the adventitious carbon on titanium substrate). In contrast, the thulium films show unusually strong high-BE components, likely indicating formates or carbonates from the deposition in addition to the adventitious carbon layer.\\
The O$_{1s}$ spectra (Fig. \ref{XPS05}b, \ref{XPS1}b) include a Ti-related oxide peak at $\approx$533~\si{\electronvolt} (indicated by the red line), and broader peaks at $\approx$533 and 537~\si{\electronvolt}. The high-BE peak at $\approx$537~\si{\electronvolt} decreases with \ce{CO2} content and increase with water content, possibly linked to oxidic species (\ce{Tm2O3}) while the peak at $\approx$530~\si{\electronvolt} increases in intensity with increasing \ce{CO2}-content, possibly linked to carbonates/formates.\\
The Tm$_{4d}$ region (Fig. \ref{XPS05}c, \ref{XPS1}c) shows a main peak between 181.4 and 185~\si{\electronvolt}, with a low-BE shoulder ($\approx$178~\si{\electronvolt}) more pronounced at depositions with high \ce{CO2}. This may reflect different thulium species or uncharged deposit regions due to the influence of the cracked thin film and possible increasing conductivity. 
XPS trends align with IR/Raman data for terbium films: oxide signals diminish while carbonate signals rise with increasing \ce{CO2}. In the 0.5~vol-\% \ce{D2O} series, high-BE C$_{1s}$ peaks (5.6–6.3~\si{\electronvolt}) appear, consistent with carbonate formation. These films, though prone to charging, indicate \ce{CO2}-induced preference for carbonates over oxides.\\

\section{Conclusion and Outlook}
\ce{CO2} and water greatly affect the layers produced by the Molecular Plating technique, emphasizing the importance of controlling solvent purity. Water is essential in the reaction solution for Molecular Plating, but excessive water concentration (>1 vol.-\%) leads to a deterioration in thin film quality. The presence of \ce{CO2} consistently degrades the thin film quality by increasing carbonate content.\\
No conclusions can be drawn from the IR and Raman spectra about the absolute quantity of the different species and their relative concentrations to each other. Only trends between the spectra can be observed. The Raman and IR spectra show clearly the presence of hydroxides and oxides. This also fits to the theory that the Molecular Plating technique is based on hydroxide precipitation \cite{Hansen1959}.\\
The IR and Raman spectra clearly indicate the presence of carbonates and formates besides hydroxides and oxides in the thin films. These formates are generated and deposited through \ce{CO2} reduction under basic conditions at the cathode.\\
Moreover, the spectra reveal that as the \ce{CO2} content in the solutions increases, the amount of carbonate rises while the relative levels of oxide and hydroxide species decrease, indicating that the reactions forming these species competes with each other. The XPS spectra further support the trends that the formation of carbonates and oxides compete with each other.\\
The obtained XPS spectra are difficult to interpret, though. Charging effects seem to affect different parts of the surface differently, leading to somewhat arbitrary shifts and fragmentation of the peaks. Due to the complex composition of the samples and the charging effects an extensive peak fitting does not seem advisable.\\
Thinner films might be less affected by charging, since the layer thickness influences its insolation and therefore the charging effects. So by using relatively thin films of maybe <100~\si[per-mode=fraction]{\micro\gram\per\square\centi\meter} that also seem to have a more homogeneous surface without major mud cracking some clarification may be obtained from XPS. On the other hand such very thin films may not represent the full reality of the final targets actually in use for the super heavy-element production.\\
In summary, the solvent quality significantly influences the success of the deposition process and requires careful control. Preserving high solvent quality can be achieved by storing them in an inert atmosphere, such as a glove box.\\
In the future, these discoveries can enhance the reliability and effectiveness of Molecular Plating. By deepening our understanding of the chemical composition, we can improve the technique itself, as well as create novel methods derived from these insights. These advancements aid the production of more effective targets for ion beam experiments and other nuclear physical experiments.\\

\section{Acknowledgement}
Our sincere thanks go the staff of the mechanical workshop at the research reactor TRIGA Mainz for their local support. We would also like to thank Constantin Haese from the Max-Planck-Institut f{\"u}r Polymerforschung Mainz who carried out NMR measurements for us. We acknowledge funding from the German Federal Ministry for Research and Education (project 05P21UMFN2). The experimental data used in this research were generated through access to the ActUsLab/PAMEC under the Framework of access to the Joint Research Centre Physical Research Infrastructures of the European Commission (Project Targets-SHE2, Research Infrastructure Access Agreement N\textsuperscript{o}~36107/01).

\section{References}
\bibliography{main}

\end{document}